\documentclass[aps,twocolumn,superscriptaddress,showpacs,showkeys,preprintnumbers,amsmath,amssymb]{revtex4}
\usepackage{txfonts}
\usepackage{dcolumn}
\usepackage{mathrsfs}% Align table columns on decimal point
\usepackage{bm}% bold math
\usepackage{amsmath,amssymb,epsfig,float,graphics}

\begin{document}

\title{ Does relativistic motion always degrade quantum Fisher information? }
\author{Xiaobao Liu}
\affiliation{Department of physics and electrical engineering, Liupanshui Normal University, Liupanshui 553004, Guizhou, China}

\author{Jiliang Jing\footnote{Corresponding author, Email: jljing@hunn.edu.cn}}
\affiliation{Department of
Physics, Key Laboratory of Low Dimensional Quantum Structures and
Quantum Control of Ministry of Education, and Synergetic Innovation
Center for Quantum Effects and Applications, Hunan Normal
University, Changsha, Hunan 410081, P. R. China}

\author{Zehua Tian \footnote{Corresponding author, Email:tianzh@ustc.edu.cn}}
\affiliation{ CAS Key Laboratory of Microscale Magnetic Resonance and Department of Modern Physics, University of Science and Technology of China, Hefei 230026, China}

\affiliation{Key Laboratory for Research in Galaxies and Cosmology,
Chinese Academy of Science, 96 JinZhai Road, Hefei 230026, Anhui, China}

\affiliation{Hefei National Laboratory for Physical Sciences at the Microscale, University of Science and Technology of China, Hefei 230026, China}

\affiliation{Synergetic Innovation Center of Quantum Information and Quantum Physics, University of Science and Technology of China, Hefei 230026, China
}

\author{Weiping Yao \footnote{Corresponding author, Email:yao11a@126.com}}
\affiliation{Department of physics and electrical engineering, Liupanshui Normal University, Liupanshui 553004, Guizhou, China}

\begin{abstract}
We investigate the ultimate estimation precision, characterized by the quantum Fisher information, of a two-level atom as a detector which is coupled to massless scalar field in the Minkowski vacuum.
It has been shown that for an inertial detector moving with a constant velocity, its quantum Fisher information is completely unaffected by the
velocity, however, it still decays over time due to the decoherence caused by the interaction between the atom and the field.  In addition, for a uniformly accelerated detector ($w=0$) moving along spatially straight line, the accelerated motion will reduce the quantum Fisher information in the estimation of state parameters. However, when the detector trajectory is generated by a combination of the linear accelerated motion and a component of the four-velocity $w=dy/d\tau$, we find quite unlike the previous results that, for the non-relativistic case $(w\ll1)$, the acceleration could degrade the quantum Fisher information, while the four-velocity component will suppress the degradation of the quantum Fisher information, and thus could enhance the precision of parameters estimation. Furthermore, in the case for ultra-relativistic velocities $(w\rightarrow\infty)$, although the detector still interacts with the environment, it behaves as if it were a closed system as a consequence of relativity correction associated to the velocity, and the quantum Fisher information in this case can be shield from the effect of the external environment, and thus from the relativistic motion.

\end{abstract}

\maketitle

%%%%%%%%%%%%%%%%%%%%%%%%%%%%%%%%%%%%%%%%%%%%%%%
\section{Introduction}
Quantum Fisher information (QFI)~\cite{Braunstein1994,Braunstein1996,Petz2011}  has attracted much interest since it is, not only of great significance in quantum estimation theory and quantum information theory~\cite{Petz1996,Petz2009,Ma2009,Rivas2010,Sun2010,Chaves2012,Ma2011+,Hyllus2012,Toth2012}, but also strongly related to rapid progress in quantum-enhanced metrology~\cite{Giovannetti2004,Giovannetti2006,Giovannetti2011}.
Indeed, in the field of quantum metrology, the QFI acted as a crucial measure of information content of quantum state, which has already played a significant role in quantum statistical inference for its inextricable relationship with Cram\'{e}r-Rao inequality, namely: the lower bound of the estimation error is characterized by the Cram\'{e}r-Rao bound which is inversely proportional to QFI~\cite{Helstrom1976,Holevo1982,bner1993}. With different models of the probe systems and different parameters to be
estimated, QFI has been applied in various quantum information processing tasks, such as measurements of non-Markovianity~\cite{Lu2010}, entanglement detection~\cite{Li2013}, qubit thermometry~\cite{Brunelli2011,Brunelli2012}, as well as relativistic parameters estimation \cite{Bruschi2014,Tian2015,Wang2014,Zhao2020}, and so on.
However, a realistic quantum system will unavoidably suffer from the quantum decoherence, due to the interaction between the system and its surrounding environment, which results in the QFI attenuation and thus the estimation
precision degradation~\cite{Huelga1997,Rosenkranz2009,Escher2011,Chin2012,Chaves2013,Ma2011,Zhong2013}. Moreover, there has been several works to investigate the degradation of the QFI caused by the effects of relativistic motion~\cite{Tian2015,Wang2014,Yao2014,Yang2018}, or the curvature of curved spacetime~\cite{Yang2020}.
In this regard, how to inhibit the the attenuation of QFI becomes the key problem to be solved.

Quite recently, the field of relativistic quantum information has emerged as an active research program, connecting concepts from gravitational physics and quantum information science.  In this respect, many investigations were implemented on the detection of relativistic effects~\cite{Aspachs2010,Ahmadi2014,2Ahmadi2014,Wang2015,sabin2014,Doukas2014,Doukasarxiv,Liu2018,Liu2019}. We also quote Refs.~\cite{Tian2015,Wang2014} which presented a heuristic scenario to estimate the Unruh effect of the uniformly linear accelerated detector coupled with scalar field within the framework of open quantum systems, have given manifest evidence that the accelerated motion will degrade  the precision of parameters estimation.
On the other hand, there existed several important works that were developed to study the detector moving along different trajectories.
For instance, it was shown in Refs.~\cite{Bell1983,Bell1987,Unruh1998,Leinaasa2002,Rogers1988,Levin1993,Davies1996,Akhmedov2007} that the detector with spatially circular trajectories was adopted to simulate the Unruh effect, particularly with relation to polarization effects of electrons in storage rings and for electrons circulating in a cavity.  Another example is related to investigations of a rotating detector coupled to a field governed by a nonlinear dispersion relation which was considered in Ref.~\cite{Gutti2011}. Moreover, in the accelerated reference frames, the authors in Ref.~\cite{Korsbakkena2004} explored the generalized Unruh effect of a detector having planar motion that include rotation in addition to acceleration. See also the reference \cite{Barbado2012} for discussions on the subject of an Unruh-DeWitt detector moving with time-dependent acceleration along a one-dimensional trajectory to analyze its response function. Currently a special case of a stationary trajectory was considered in Ref.~\cite{Abdolrahimi2014}. In that work, the author has shown some interesting results, when the detector moves along an unbounded spatial trajectory in a two-dimensional spatial plane with constant independent magnitudes of both the four-acceleration and of a timelike proper time derivative of four-acceleration, and having a constant component of four-velocity. Consequently, if we consider the relativistic motion of Unruh-DeWitt detector is a superposition of both a linear accelerated motion and a component of the four-velocity, whether can suppress the degradation of the QFI or not?

In this paper, we will study the performance of QFI regarding the estimation of parameters for a two-level system as the detector coupled to massless scalar field. Here we consider this detector moving in Minkowski spacetime along an unbounded spatial trajectory in a two-dimensional spatial plane with
constant independent magnitudes of both the four-acceleration and of a timelike proper time derivative of the four-acceleration, which was shown in Ref.~\cite{Abdolrahimi2014}. In such a reference, in a Fermi-Walker frame moving with the detector, the direction of the acceleration rotates at a constant rate around a great circle.
Our analytical results demonstrate that in the non-relativistic limit, the four-velocity component will suppress the attenuation of the QFI, which implies that the precision of parameters estimation of the Unruh-DeWitt detector moving along such a trajectory can be enhanced. What is more, in the ultra-relativistic limit,
the QFI may even be shield from the effects of the external environment and detector's motion, as if this detector were a closed system.

Our paper is structured as follows. In Sec. \ref{sectionII}, we introduce the QFI and the dynamic evolution of detector coupling with massless scalar field.
In  Sec. \ref{sectionIII}, we study the QFI in the parameters estimation in two situations: for non-relativistic and ultra-relativistic velocities. Finally, in Sec. \ref{sectionIV}, we give our conclusions and discussions.

Throughout the whole paper we employ natural units $c = \hbar = 1$. Relevant constants are restored when needed for the sake of clarity.

%%%%%%%%%%%%%%%%%%%%%%%%%%%%%%%%%%%%%%%%%%%%%%%%%%%%%%%%%%%%%%%%%%%%%%%%%%%%%%

\section{quantum Fisher information and dynamic evolution of a two-level system coupled with scalar fields}\label{sectionII}
In quantum metrology, any given quantum state $\rho(X)$ characterized by the unknown parameter $X$ can be inferred from a set of measurements on the state.
The measurements usually modeled mathematically by a set of positive operator-valued measures (POVM), whose elements, $\{\Pi_{i}\}$, saturate to
$\sum_{i}\Pi_{i}\Pi_{i}^{\dag}=1$. Through the optimization of the measurements and the estimator, an ultimate bound to precision of the unknown parameter estimation satisfies the quantum Cramer-Rao inequality~\cite{Braunstein1994}
\begin{eqnarray}
\rm{Var} \;(X)\geq \frac{1}{M F_{X}}\;,
\end{eqnarray}
where $M$ represents the number of measurements, and $F_{X}=\rm{Tr}[\rho(X)L^{2}]$ is the QFI. Here, $L$ denotes the symmetric logarithmic derivative satisfying the partial differential equation
\begin{eqnarray}
\frac{\partial\rho(X)}{\partial X}=\frac{L\rho(X)+\rho(X)L}{2}\;.
\end{eqnarray}
For a two-level quantum system, the reduced density matrix of the system can be expressed in the Bloch sphere representation as
\begin{eqnarray}\label{PM}
\rho(\tau)=\frac{1}{2}\left(\hbox{I}+\boldsymbol{\omega}(\tau)\cdot\boldsymbol{\sigma}\right),
\end{eqnarray}
where $\boldsymbol{\omega}=(\omega_1,\omega_2,\omega_3)$ represents the Bloch vector, and $\boldsymbol{\sigma}=(\sigma_1,\sigma_2,\sigma_3)$ denotes the Pauli matrices.
As shown in Ref~\cite{Zhong2013}, $F_{X}$ can be described in the simple form as follow:
\begin{eqnarray}\label{QFI}
F_{X}=
\begin{cases}
\;|\partial_{X}\boldsymbol{\omega}|^2+\frac{\left(\boldsymbol{\omega}\partial_{X}\boldsymbol{\omega}\right)^2}{1-|\boldsymbol{\omega}|^2},\;
\;\;\;|\boldsymbol{\omega}| < 1\;,\\
\;|\partial_{X}\boldsymbol{\omega}|^2,\;\;\;\;\;\;\;\;\;\;\;\;\;\;\;\;\;\;|\boldsymbol{\omega}| = 1\;.
\end{cases}
\end{eqnarray}

In quantum sense, any system should be regarded as an open system due to the interaction between the system and its surrounding environments.
Therefore, let us study a two-level detector interacting with massless scalar field and, in this regard, the total Hamiltonian of the detector-field system can be described as
\begin{eqnarray}\label{Hamiltonian0}
H=H_s+H_{\Phi(x)}+H_I,
\end{eqnarray}
where $H_s=\frac{1}{2}\omega_0\sigma_z$ is the Hamiltonian of the detector with $\omega_0$ and $\sigma_z$ being the energy-level spacing of the atom and Pauli matrix, respectively, $H_{\Phi(x)}$ represents the Hamiltonian of scalar field and $H_I$ denotes their interaction Hamiltonian.
We assume that the coupling between the detector and the massless scalar field is of the form,
\begin{eqnarray}
H_I=\mu(\sigma_+ + \sigma_-)\Phi(x(\tau))\;,
\end{eqnarray}
where $\mu$ is the coupling constant that we assume to be small, $\sigma_+$ ($\sigma_-$) is the rasing (lowering) operator of the detector, and $\Phi(x(\tau))$ corresponds to the scalar field operator with $\tau$ being the detector's proper time.
Note that this interaction is the analogy to the electric dipole interaction. Specifically, $H_I=\mu'\hat{m}\Phi(x(\tau))=\sum_{i,j=0}^1|i\rangle\langle i|\mu' \hat{m} |j\rangle\langle j| \Phi(x(\tau))=\mu(\sigma_++\sigma_-)\Phi(x(\tau))$, where we have used $\sum_i |i\rangle\langle i|=I$ and
$\langle i|\mu \hat{m} |i\rangle=0$.

We assume the scalar field in vacuum state $|0\rangle$, which is defined by $a_\mathbf{k}|0\rangle=0$ for all $\mathbf{k}$.
At the beginning, the total density matrix of the detector-field system can be written as $\rho_{tot}=\rho_s(0)\otimes|0\rangle\langle0|$, in which $\rho_s(0)$ is the initial reduced density matrix of the detector.
For the whole system, its equation of motion in the interaction picture is given by
\begin{eqnarray}\label{whole}
\frac{\partial\rho_{tot}(\tau)}{\partial\tau}&=&-i[H_I(\tau),\rho_{tot}(\tau)]\;.
\end{eqnarray}
We have $\rho_{tot}(\tau)=\rho_{tot}(0)-i\int^\tau_0ds[H_I(s),\rho_{tot}(s)]$. Substitute this solution back into
Eq. \eqref{whole} and take the partial trace:
\begin{eqnarray}\label{equation1}
\frac{\partial\rho_s(\tau)}{\partial\tau}=-\int_0^\tau dsTr_B[H_I(\tau),[H_I(s),\rho_{tot}(s)]]\;,
\end{eqnarray}
where we have used $Tr_B[H_I(\tau),\rho_{tot}(0)]=0$. We now make our first approximation. For a sufficiently large bath that is in particular much larger than the system, it is reasonable to assume that while the system undergoes non-trivial evolution, the bath remains unaffected, and hence that the state of the composite system at any time is uncorrelated, i.e., \begin{eqnarray}
\rho_{tot}(s)\approx\rho_s(s)\otimes\rho_B\;.
\end{eqnarray}
This is the so called Born approximation \cite{Breuer2002}. Let us change the variables to $\tau^\prime=\tau-s$, so that $\int^\tau_0ds=-\int^0_\tau\,d\tau^\prime=\int^\tau_0d\tau^\prime$, and Eq. \eqref{equation1} can be rewritten as
\begin{eqnarray}\label{equation2}
\frac{\partial\rho_s(\tau)}{\partial\tau}&=&-\int_0^\tau d\tau^\prime\,Tr_B[H_I(\tau),[H_I(\tau-\tau^\prime),\rho_s(\tau-\tau^\prime)\otimes\rho_B]].\nonumber\\
\end{eqnarray}
We then introduce the so called Markov approximation \cite{Breuer2002}. It states that the bath has a very short correlation time $\tau_B$. If
$\tau\gg\tau_B$, we can replace $\rho_s(\tau-\tau^\prime)$ by $\rho_s(\tau)$, since the short ``memory" of the bath correlation function  causes it to keep track of events only within the short period $[0,\tau_B]$. Under this approximations, we have
\begin{eqnarray}\label{equation3}
\frac{\partial\rho_s(\tau)}{\partial\tau}=-\int_0^\tau d\tau^\prime\,Tr_B[H_I(\tau),[H_I(\tau-\tau^\prime),\rho_s(\tau)\otimes\rho_B]]\;.\nonumber\\
\end{eqnarray}
Moreover, for the same reason (correlation function negligible for $\tau^\prime\gg\tau_B$ \cite{Breuer2002}) we can extend the upper limit of the integral to infinity without changing the value of the integral. Therefore,
\begin{eqnarray}\label{equation4}
\frac{\partial\rho_s(\tau)}{\partial\tau}=-\int_0^\infty d\tau^\prime\,Tr_B[H_I(\tau),[H_I(\tau-\tau^\prime),\rho_s(\tau)\otimes\rho_B]]\;.\nonumber\\
\end{eqnarray}
Substitute the interaction Hamiltonian $H_I(\tau)=\mu (\sigma_+e^{i\omega_0\tau} + \sigma_-e^{-i\omega_0\tau}) \Phi(x(\tau))$ into Eq. \eqref{equation4} and after long but straightforward calculations, we can derive finally the master equation in the Kossakowski-Lindblad form~\cite{Lindblad1,Lindblad2,Benatti1}
\begin{eqnarray}\label{Lindblad equation}
\frac{\partial\rho_s(\tau)}{\partial\tau}&=&-i[H_{eff},\rho_s(\tau)]+\sum^3_{j=1}[2L_j\rho_s L^\dagger_j-L^\dagger_jL_j\rho_s-\rho_s L^\dagger_jL_j],\nonumber\\
\end{eqnarray}
where
\begin{eqnarray}\label{Effective H}
H_{eff}=\frac{1}{2}\Omega\sigma_z=\frac{1}{2}\{\omega_0+\mu^2\mathrm{Im}(\Gamma_++\Gamma_-)\}\sigma_z
\end{eqnarray}
is the effective Hamiltonian in which $\Omega$ denotes the effective energy level-spacing of the detector with a correction term $\mu^2\mathrm{Im}(\Gamma_++\Gamma_-)$ being the Lamb shift.  Note that the Lamb shift can be neglected because it is far less than $\omega_0$, i.e., $\Omega\approx\omega_0$.
We have defined
\begin{eqnarray}\label{gamma}
\gamma_{\pm}&=&2\mu^2{\rm Re}\Gamma_\pm=\mu^2\int^{+\infty}_{-\infty}e^{\mp i\omega_0\triangle\tau}G^+(\triangle\tau- i\epsilon)d\triangle\tau\;,\nonumber\\
\gamma_{z}&=&0\;,\nonumber\\
L_1&=&\sqrt{\frac{\gamma_-}{2}}\sigma_-\;,L_2=\sqrt{\frac{\gamma_+}{2}}\sigma_+\;,L_3=\sqrt{\frac{\gamma_z}{2}}\sigma_z\;,
\end{eqnarray}
where $\triangle\tau=\tau-\tau'$. Here, $G^+(\triangle\tau)$ is given by $G^+(x-x')=\langle0|\Phi(x(\tau))\Phi(x(\tau'))|0\rangle$
being the two-point correlation function, which for massless scalar field reads~\cite{Birrell}
\begin{eqnarray}\label{Wightman}
&&G^+(x-x')\nonumber\\
&&=\frac{1}{4\pi^2[(x-x')^2+(y-y')^2+(z-z')^2-(t-t'-i\epsilon)^2]}\;,\nonumber\\
\end{eqnarray}
where $\epsilon$ is an infinitesimal constant.

We take the initial state of the detector as
\begin{eqnarray}\label{initial state}
|\psi(0)\rangle= \sin\frac{\theta}{2}|0\rangle + e^{-i\phi}\cos\frac{\theta}{2}|1\rangle\;,
\end{eqnarray}
where $\theta$ and $\phi$ denote the initial weight parameter and phase parameter, and $|0\rangle$, $|1\rangle$ are the ground state and excited state of the detector, respectively.
By substituting the density matrix $\rho(\tau)$ Eq.~(\ref{PM}) into the master equation (\ref{Lindblad equation}), the time dependent state parameters $\boldsymbol{\omega}(\tau)$ in terms of the proper time $\tau$, after a series of calculations, are found to be
\begin{eqnarray}\label{PM1}
\omega_1(\tau)&=&\omega_1(0)\cos(\Omega\tau) e^{-\frac{1}{2}A\tau}-\omega_2(0)\sin(\Omega\tau) e^{-\frac{1}{2}A\tau}\;,\nonumber\\
\omega_2(\tau)&=&\omega_1(0)\sin(\Omega\tau) e^{-\frac{1}{2}A\tau}+\omega_2(0)\cos(\Omega\tau) e^{-\frac{1}{2}A\tau}\;,\nonumber\\
\omega_3(\tau)&=&\omega_3(0) e^{-A\tau}+\frac{B}{A}(1-e^{-A\tau})\;,
\end{eqnarray}
where $A=\gamma_+ +\gamma_-$, $B=\gamma_+ -\gamma_-$, and $\lim_{\tau\rightarrow0}\omega_i(\tau)=\omega_i(0)$.
Substituting the initial state~(\ref{initial state}) into Eq.~(\ref{PM1}), the general analytic solution of the evolution of two-level system then can be written as
\begin{eqnarray}\label{final state}
\omega_1(\tau)&=&\sin\theta \cos(\Omega\tau+\phi) e^{-\frac{1}{2}A\tau}\;,\nonumber\\
\omega_2(\tau)&=&\sin\theta \sin(\Omega\tau+\phi) e^{-\frac{1}{2}A\tau}\;,\nonumber\\
\omega_3(\tau)&=&\cos\theta e^{-A\tau}+\frac{B}{A}(1-e^{-A\tau})\;.
\end{eqnarray}

%%%%%%%%%%%%%%%%%%%%%%%%%%%%%%%%%%%%%%%%%%%%%%%%%%%%%%%%%%%%%%%%%%%%%%%%%%%%%%

\section{Relativistic motion affects on parameters estimation} \label{sectionIII}
Now let us first calculate the QFI for the parameter estimation of the detector, moving along a spatially straight line with constant four-velocity component $w$, whose spacetime coordinates are given by~\cite{Abdolrahimi2014}
\begin{eqnarray}\label{trajectory1}
t(\tau)=\sqrt{1+w^2}\tau\;,\;\;x(\tau)=0\;,\;\;y(\tau)=w\tau\;,\;\;z(\tau)=0\;.
\end{eqnarray}
Submitting Eqs.~(\ref{Wightman}) and (\ref{trajectory1}) into Eq.~(\ref{gamma}), the $A$ and $B$ can be calculated easily as
\begin{eqnarray}
A=\gamma_0\;,\;\;\;B=-\gamma_0\;,
\end{eqnarray}
where $\gamma_0=\frac{\mu^2\omega_0}{2\pi}$ is the spontaneous emission rate.
Thus, the Bloch vector of detector's state evolves with proper time can be obtained as
\begin{eqnarray}
\omega_1(\tau)&=&\sin\theta \cos(\Omega\tau+\phi) e^{-\frac{1}{2}\gamma_0\tau}\;,\nonumber\\
\omega_2(\tau)&=&\sin\theta \sin(\Omega\tau+\phi) e^{-\frac{1}{2}\gamma_0\tau}\;,\nonumber\\
\omega_3(\tau)&=&\cos\theta e^{-\gamma_0\tau}-(1-e^{-\gamma_0\tau})\;.
\end{eqnarray}
As a result, the QFI of the initial weight $\theta$ and phase parameter $\phi$ become $F_\theta=e^{-\gamma_0\tau}$ and $F_\phi=\sin^2\theta e^{-\gamma_0\tau}$. It implies that the QFI of both weight and phase parameters decreases exponentially with time, due to the decoherence caused by the interaction between the detector and the massless scalar field. However, the QFI is completely unaffected under the effect of the four-velocity component.

However, in this paper we may wonder how the detector motion is generated by a combination of the linear accelerated motion and a component of the four-velocity affects on the performance of QFI of parameters estimation.
Now we consider the detector moving in flat spacetime along an unbounded spatial trajectory in a two-dimensional spatial plane with a constant square of magnitude of four-acceleration $a_\mu a^\mu=a^2$, and constant magnitudes of a timelike proper time derivative of four-accelration $(da_{\mu}/d\tau)(da^{\mu}/d\tau)$, which has a constant component of the four-velocity $w=dy/d\tau=const$. Here, $a^\mu=d^2x^{\mu}/d\tau^2$. As a result, the spacetime coordinates of detector are described by~\cite{Abdolrahimi2014,Letaw1981}
\begin{eqnarray}\label{trajectory}
&&t(\tau)=\frac{a}{\alpha^2}\sinh\alpha\tau\;,\;\;\;x(\tau)=\frac{a}{\alpha^2}\cosh\alpha\tau\;,\nonumber\\
&&y(\tau)=w\tau\;,\;\;\;\;\;\;\;\;\;\;\;\;\;\;\;z(\tau)=0\;,
\end{eqnarray}
where $\alpha=\frac{a}{\sqrt{1+w^2}}>0$.
Applying the trajectory of detector (\ref{trajectory}) into Eq.~(\ref{Wightman}), the two-point correlation function for the massless scalar field is given by
\begin{eqnarray}\label{correlation}
G^+(x-x')=-\frac{\alpha^4}{16\pi^2}\bigg[\sinh^2\bigg(\frac{\alpha\Delta\tau}{2}-\frac{i\epsilon\alpha^2}{a}\bigg)-\frac{w^2\alpha^4}{4a^2}\Delta\tau^2\bigg]^{-1}.\nonumber\\
\end{eqnarray}
Note that for $w=0$, we have $\alpha=a$ and the two-point correlation function in Eq.~(\ref{correlation}) recovers to that of a detector moving along a spatially straight line along the $x$ direction with constant magnitude of the four-acceleration (spatially one-dimensional)~\cite{Tian2015,Wang2014}, as expected, which is
\begin{eqnarray}
G^{+(0)}(\Delta\tau)=-\frac{a^2}{16\pi^2}\bigg[\sinh^2\bigg(\frac{a\Delta\tau}{2}-i\epsilon a\bigg)\bigg]^{-1}.
\end{eqnarray}
In the following, we are interested in investigating the QFI of this detector moving along such a trajectory in Eq.~(\ref{trajectory}) in two situations: in the non-relativistic and ultra-relativistic limit, respectively.

%%%%%%%%%%%%%%%%%%%%%%%%%%%%%%%%%%%%%%%%%%%%%%%%%%%%%%%%%%%%%%%%%%%%%%%%%%%%%%

\subsection{In the case for non-relativistic velocities}
Let us consider the case for the Unruh-DeWitt detector moving along such a trajectory in Eq.~(\ref{trajectory}) in the non-relativistic limit, i.e., $w\ll 1$, the two-point correlation function for the massless scalar field in Eq.~(\ref{correlation}) up to second order in $w^2$ which is
\begin{eqnarray}\label{noncorrelation}
G^+(x-x')&=&(1-2w^2)G^{+(0)}(\Delta\tau)\nonumber\\
&&-\frac{a^2}{16\pi^2}\bigg[\sinh(a\Delta\tau-2i\epsilon a)\bigg(\frac{a\Delta\tau}{4}-i\epsilon a\bigg)+\frac{a^2\Delta\tau^2}{4}\bigg]\nonumber\\
&&\times\bigg[\sinh^4\bigg(\frac{a\Delta\tau}{2}-\frac{i\epsilon\alpha^2}{a}\bigg)\bigg]^{-1} w^2.
\end{eqnarray}
Applying Eq.~(\ref{noncorrelation}) to Eq.~(\ref{gamma}), and through the contour integral, we can obtain
\begin{eqnarray}
\gamma_{\pm}=2\mu^2\bigg[\mathcal{G}^{0}(\pm\omega_0)+\mathcal{G}^{1}(\pm\omega_0)w^2\bigg]
\end{eqnarray}
with
\begin{eqnarray}
&&\mathcal{G}^{0}(\omega_0)=\frac{\omega_0}{2\pi}\bigg(\frac{1}{e^{2\pi\omega_0/a}-1}\bigg)\;,\nonumber\\
&&\mathcal{G}^{0}(-\omega_0)=\frac{\omega_0}{2\pi}\bigg(\frac{e^{2\pi\omega_0/a}}{e^{2\pi\omega_0/a}-1}\bigg)\;,
\nonumber\\
&&\mathcal{G}^{1}(\pm\omega_0)=-f(a)\;,
\end{eqnarray}
where $f(a)=\frac{a e^{2\pi\omega_0/a}}{6[e^{2\pi\omega_0/a}-1]^2}
\bigg[2+\frac{9\omega_0^2}{a^2}-\frac{2\pi\omega_0}{a}\bigg(1+\frac{\omega_0^2}{a^2}\bigg)\frac{e^{2\pi\omega_0/a}+1}{e^{2\pi\omega_0/a}-1}\bigg]$.
Therefore, we have $A$ and $B$ in Eq.~(\ref{final state}) which are
\begin{eqnarray}\label{Veffects}
&&A=\gamma_0\bigg[\frac{e^{2\pi\omega_0/a}+1}{e^{2\pi\omega_0/a}-1}-\frac{4\pi}{\omega_0}f(a)w^2\bigg]\;,\nonumber\\
&&B=-\gamma_0\;.
\end{eqnarray}

%%%%%%%%%%%%%%%%%%%%%%%%%%%%%%%%%%%%%%%%%%%%%%%%%%%%%%%%%%%%%%%%%%%%%%%%%%%%%%

\subsubsection{Relativistic motion affects on the precision in the estimation of phase parameter $\phi$}
We discuss the relativistic motion of the detector in Eq.~(\ref{trajectory}) how to affect the precision in the estimation of phase parameter $\phi$ in the non-relativistic limit. For the sake of simplicity, in this paper we will work with dimensionless quantities by rescaling time $\tau$ and four-acceleration $a$
\begin{eqnarray}\label{dimensionless}
\tilde{\tau}\equiv \gamma_0 \tau\;, \ \ \
\tilde{a}\equiv \frac{a}{\omega_0}\;.
\end{eqnarray}
Substituting Eqs.~(\ref{final state}) and (\ref{Veffects}) into Eq.~(\ref{QFI}), one can easily obtain the detailed formula of the QFI with respect to $\phi$ as
\begin{eqnarray}\label{QFIphi}
F_{\phi}=\sin^2\theta \; e^{-h(\tilde{a})\tilde{\tau}}\;,
\end{eqnarray}
where $h(\tilde{a})=\frac{e^{2\pi/\tilde{a}}+1}{e^{2\pi/\tilde{a}}-1}-4\pi f(\tilde{a})w^2$. It is worth mentioning that $\gamma=h(\tilde{a})\gamma_0$ represents the decay rate for a two-level detector moving the trajectory~(\ref{trajectory}) with a component of the four-velocity $w=dy/d\tau=const$.
Interesting, we notice that the QFI in Eq.~(\ref{QFIphi}) is irrespective of quantum phase $\phi$, but depends on the value of initial weight parameter $\theta$, time $\tilde{\tau}$, four-acceleration $\tilde{a}$ and four-velocity component $w$.
Hereafter, for convenience, we continue to term $\tilde{\tau}$ and $\tilde{a}$ as $\tau$ and $a$, respectively, in this paper.

\begin{figure}[htbp]
\centering
\includegraphics[height=2.0in,width=3.0in]{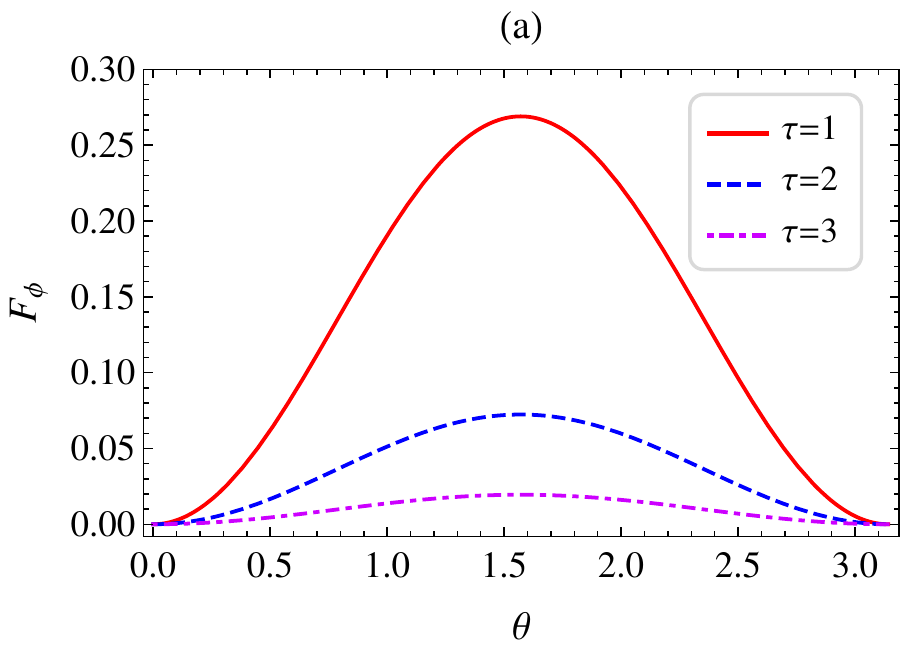}
\includegraphics[height=2.0in,width=3.0in]{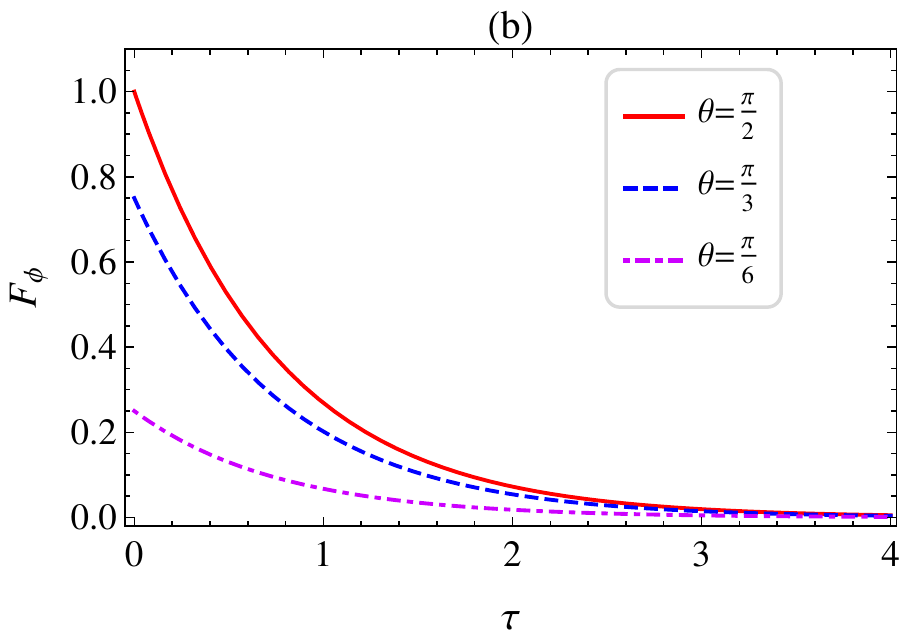}
\caption{  (a) The QFI $F_{\phi}$ as a function of the initial weight parameter $\theta$ with different effective time $\tau=1$ (solid line), $\tau=2$ (dashed line), $\tau=3$ (dot-dashed line); (b) The QFI  $F_{\phi}$ as a function of the effective time $\tau$ with $\theta=\pi/2$ (solid line), $\theta=\pi/3$ (dashed line), $\theta=\pi/6$ (dot-dashed line). Here, we take the effective four-acceleration $a=\pi$ and four-velocity component $w=0.01$.}\label{QFI1}
\end{figure}

To show the properties of the precision of the phase parameter estimation, we plot, in Fig.~\ref{QFI1}, the QFI as the function of the initial state parameter $\theta$ (effective time $\tau$) with different effective time $\tau$ (initial weight parameter $\theta$).
We are interested in finding that the maximal $F_{\phi}$ can be achieved by taking $\theta=\pi/2$, i.e., by preparing the two-level detector in the \emph{balance-weighted state} which is preferable (see figure \ref{QFI1}).
From Fig.~\ref{QFI1}~(a), we see that the QFI in the estimation of phase parameter $\phi$ is symmetric with respect to $\theta=\pi/2$. Moreover, in Fig.~\ref{QFI1}~(b), the QFI $F_{\phi}$ takes the maximum value when the quantum state is at the beginning ($\tau=0$), which implies that the precision in the estimation of phase parameter decreases with the effective evolution time, because the decoherence is caused by the interaction between the detector and massless scalar field.

\begin{figure}[htbp]
\centering
\includegraphics[height=2in,width=3in]{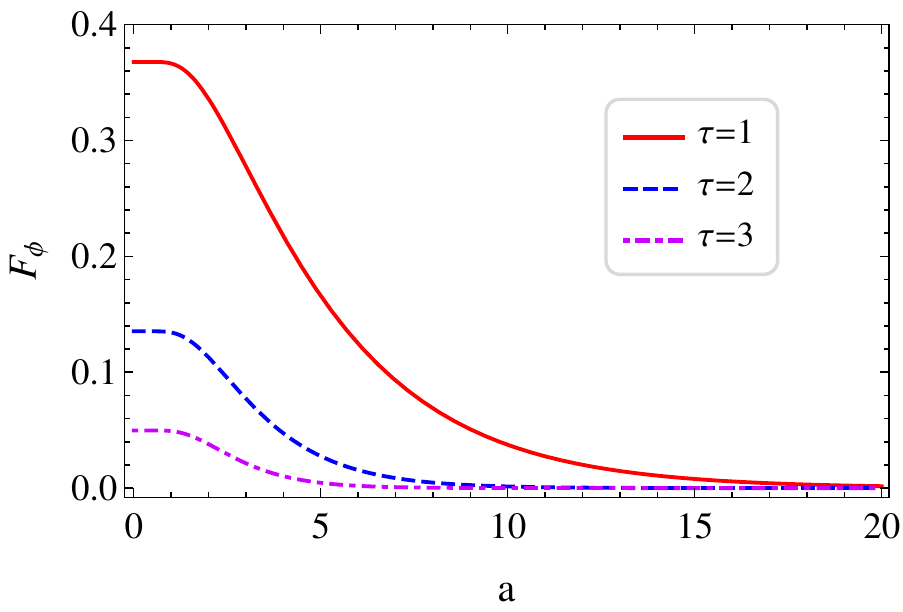}
\caption{ The QFI $F_{\phi}$ as a function of the effective four-acceleration parameter $a$ with three different effective evolution time $\tau$. We take $\theta=\pi/2$ and $w=0.01$.}\label{QFI2}
\end{figure}
As can be seen in Fig.~\ref{QFI2}, the QFI in the estimation of phase parameter $\phi$ is plotted as a function of the effective four-acceleration $a$. We observe that as the effective four-acceleration $a$ increases, the QFI $F_{\phi}$ gradually decreases and converges to zero value in the limit of infinite four-acceleration, which is reminiscent of previous results that the quantum entanglement vanishes with infinite acceleration~\cite{Fuentes}. The reason is that the larger effective four-acceleration results in the larger decay rate of the atom which is modified by the factor $[\frac{e^{2\pi/a}+1}{e^{2\pi/a}-1}-4\pi f(a)w^2]$ comparing with the spontaneous emission rate, which implies that the precision of phase parameter is an decreasing function of the four-acceleration.

\begin{figure}[htbp]
\centering
\includegraphics[height=2in,width=3in]{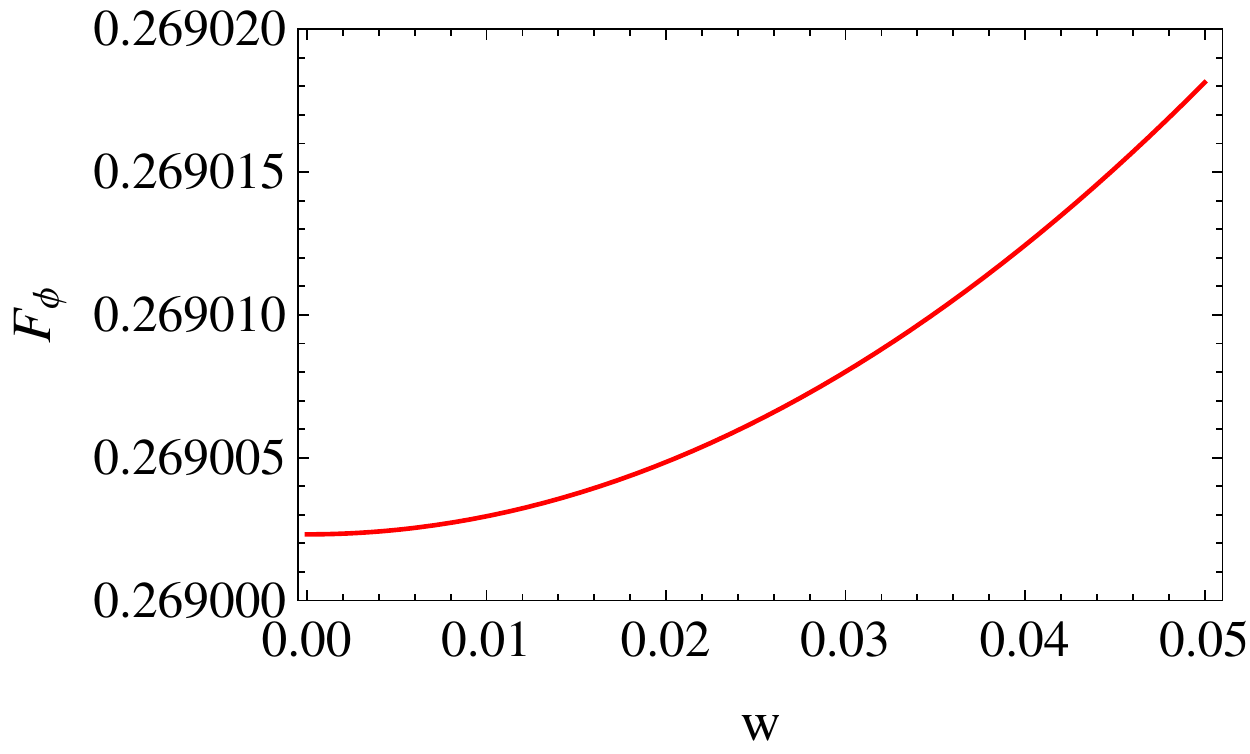}
\caption{  QFI in the estimation of phase parameter as a function of the four-velocity component $w$. By preparing the detector in the balance-weighted state $\theta=\pi/2$, the effective time $\tau=1$ and the effective four-acceleration $a=\pi$.}\label{QFI3}
\end{figure}

More remarkably, to analyze the four-velocity component for the accelerated Unruh-DeWitt detector in the non-relativistic limit how to affect the precision in the estimation of phase parameter, we show the QFI $F_{\phi}$ as a function of the four-velocity component $w$ in Fig.~\ref{QFI3}.
We are interested in noting that the higher the four-velocity component $w$, the bigger the QFI is, i.e., it is easier to achieve a given precision of phase parameter estimation. As a result, we can infer that in the non-relativistic limit, the four-velocity component can suppress the degradation of the QFI, which means that the precision in the estimation of phase parameter $\phi$ is enhanced when the detector moves along such a stationary trajectory in Eq.~(\ref{trajectory}).

%%%%%%%%%%%%%%%%%%%%%%%%%%%%%%%%%%%%%%%%%%%%%%%%%%%%%%%%%%%%%%%%%%%%%%%%%%%%%%

\subsubsection{Relativistic motion affects on the precision in the estimation of initial weight parameter $\theta$}
Then we want to examine the relativistic motion of the detector in Eq.~(\ref{trajectory}) how to affect the precision in the estimation of the initial weight parameter $\theta$ in the non-relativistic limit. With the help of Eqs.~(\ref{QFI}), (\ref{final state}) and (\ref{Veffects}), we can evaluate the QFI in terms of $\theta$ which is
\begin{eqnarray}\label{QFIweight}
F_\theta &=&e^{-h(a)\tau}\bigg\{\cos^2\theta+\sin^2\theta \;e^{-h(a) \tau}\nonumber\\
&&\times\bigg[1-\frac{(1-e^{h(a) \tau})^2[1+h(a)\cos\theta]^2}{[h(a)^2-1]e^{h(a) \tau}+[1+h(a)\cos\theta]^2}\bigg]\bigg\},
\end{eqnarray}
where $\tau$ is the effective time, and the factor $h(a)=\frac{e^{2\pi/a}+1}{e^{2\pi/a}-1}-4\pi f(a)w^2$ with $a$ being the effective four-acceleration.
Let us note that the QFI $F_\theta$ only depends on the initial weight parameter $\theta$, effective time $\tau$, effective four-acceleration $a$ and four-velocity component $w$, but is independent of the phase parameter $\phi$ of the detector.

\begin{figure}[htbp]
\centering
\includegraphics[height=2in,width=3in]{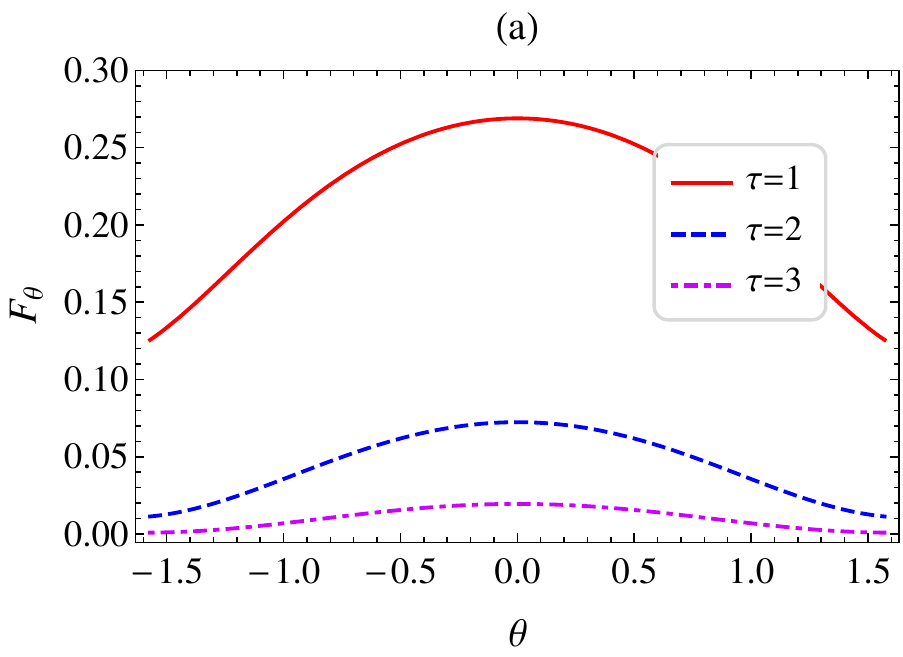}
\includegraphics[height=2in,width=3in]{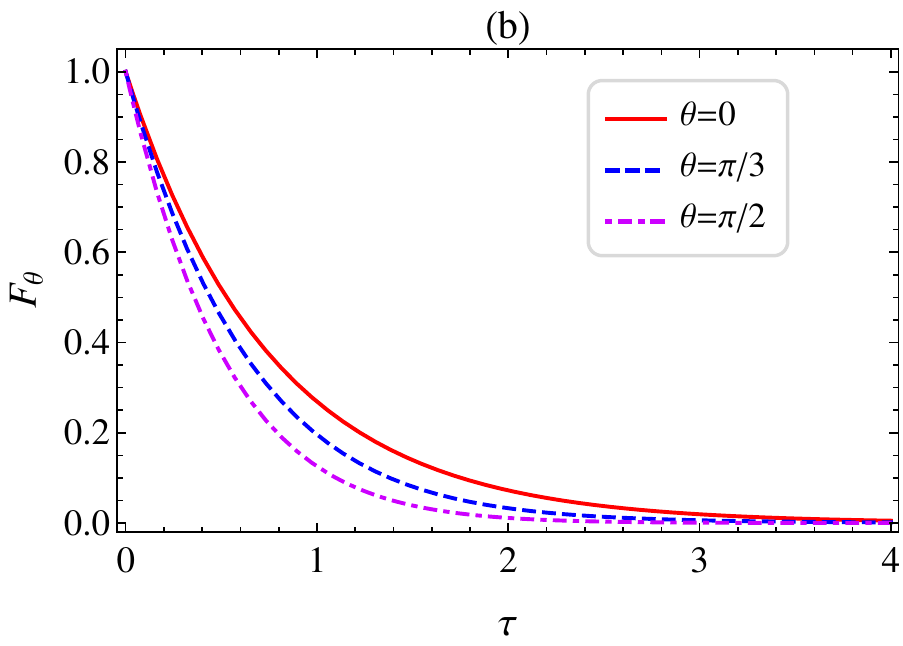}
\caption{  (a) The QFI $F_\theta$ as a function of the initial weight parameter $\theta$ with fixed values of $\tau$, i.e., $\tau=1$ (solid line), $\tau=2$ (dashed line), $\tau=3$ (dot-dashed line);
(b) The QFI  $F_\theta$ as a function of the effective time $\tau$ with fixed values of $\theta$, i.e., $\theta=0$ (solid line), $\theta=\pi/3$ (dashed line), $\theta=\pi/2$ (dot-dashed line). Here, we take the effective four-acceleration $a=\pi$ and four-velocity component $w=0.01$.}\label{QF1}
\end{figure}
Similarly, to show the behaviors of the precision in the estimation of $\theta$, in Fig.~\ref{QF1}, we plot the QFI $F_{\theta}$ as a function
of the initial weight parameter $\theta$ (effective time $\tau$) with different effective time $\tau$ (initial weight parameter $\theta$).
As we can see from Fig.~\ref{QF1}~(a),  the maximal QFI $F_{\theta}$ is obtained by taking $\theta=0$. That is, the precision in the estimation of initial weight parameter can be achieved by preparing the detector in the \emph{excited state}. In addition, we find the symmetry of the function of $F_{\theta}$ with respect to $\theta=0$. It is obvious from Fig.~\ref{QF1}~(b) that the QFI $F_{\theta}$ decreases by increasing the value of effective time $\tau$, which means that the precision in the estimation of initial weight parameter reduced by the decoherence of the detector which is caused by the interaction between the detector and
field. Moreover, in Fig.~\ref{QF1}~(b), we note that the maximal value of the QFI is obtained initially, i.e., $F_{\theta}=1$, which implies that the QFI $F_\theta$ is immune to the external environment at the beginning ($\tau=0$). This result is sharp contrast with the behavior of the QFI in the estimation of phase parameter $\phi$ shown in Fig.~\ref{QFI1}~(b).

\begin{figure}[htbp]
\centering
\includegraphics[height=2in,width=3in]{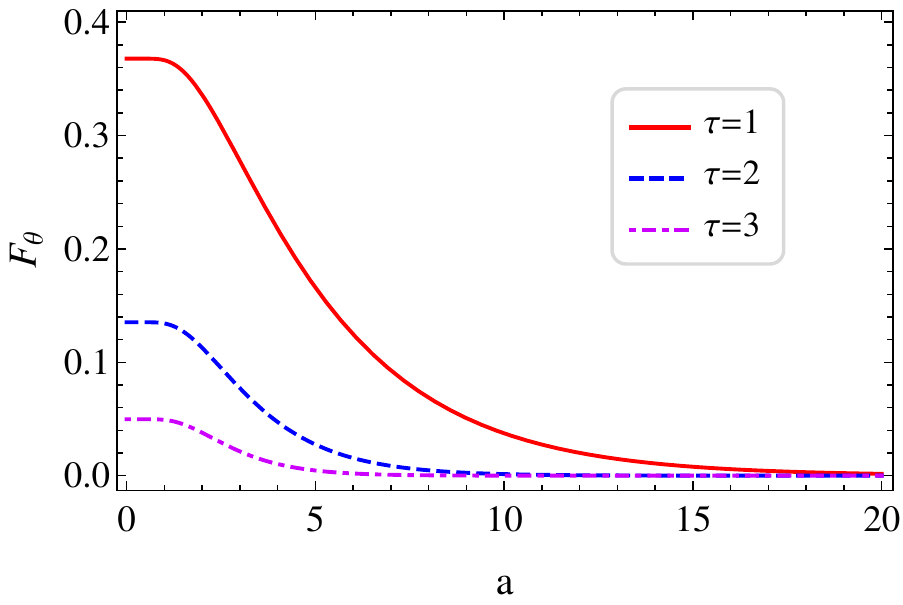}
\caption{
The QFI $F_\theta$ as a function of the effective four-acceleration parameter $a$ with the effective time $\tau=1$ (solid line), $\tau=2$ (dashed line), $\tau=3$ (dot-dashed line). We take $\theta=0$ and $w=0.01$.}\label{QF2}
\end{figure}

In Fig.~\ref{QF2}, we plot the QFI of initial weight parameter in Eq.~(\ref{QFIweight}) as a function of the effective four-acceleration parameter $a$ with different effective time $\tau$ at fixed $w=0.01$ for $\theta=0$. In a similar way, we find that as the effective four-acceleration parameter $a$ gets larger values, the QFI $F_\theta$ decreases and reduces to zero in the limit of infinite four-acceleration, which means that the precision in the estimation of initial weight parameter decreases as the effective four-acceleration increases. This is due to the fact that the larger effective four-acceleration results in a larger decay rate.

\begin{figure}[htbp]
\centering
\includegraphics[height=2in,width=3in]{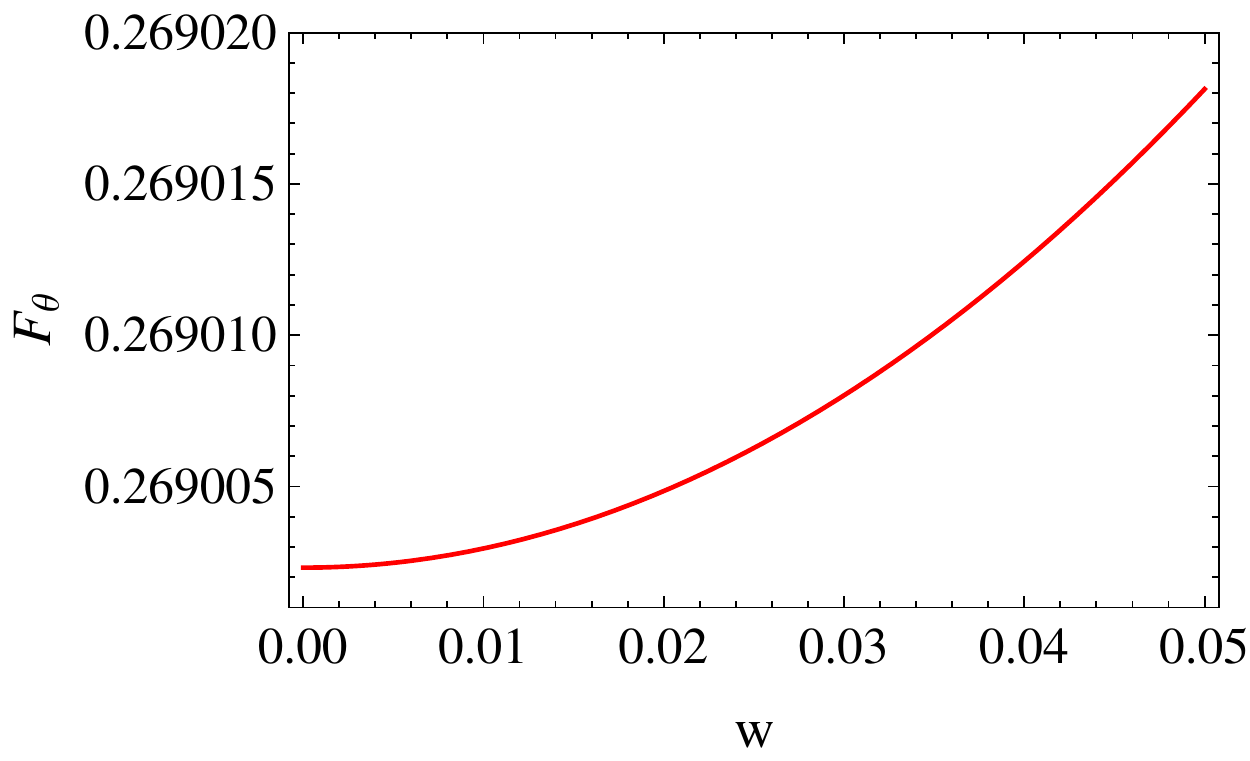}
\caption{  The QFI $F_\theta$ as a function of the four-velocity component $w$. Here, we take the detector in the exited state $\theta=0$, the effective time $\tau=1$ and the effective four-acceleration $a=\pi$.}\label{QF3}
\end{figure}

To assess the performance of the four-velocity component for the accelerated Unruh-DeWitt detector in the non-relativistic limit how to influence the precision in the estimation of initial weight parameter, figure \ref{QF3} represents the QFI $F_\theta$ as a function of the four-velocity component $w$. It is worthy noting from Fig.~\ref{QF3} that the QFI is increased as the growth of the four-velocity component $w$, which indicates that the highest precision in the estimation of initial weight parameter can be obtained for a larger four-velocity component. Thus, we argue that when the detector follows such trajectory shown in Eq.~(\ref{trajectory}) for non-relativistic velocities, the quantum estimation of initial weight parameter can be enhanced by the four-velocity component $w$, i.e., such relativistic motion of detector in the non-relativistic limit can provide us a better precision.

%%%%%%%%%%%%%%%%%%%%%%%%%%%%%%%%%%%%%%%%%%%%%%%%%%%%%%%%%%%%%%%%%%%%%%%%%%%%%%

\subsubsection{Relativistic motion affects on the precision in the estimation of parameter $\beta$}
In this section, we want to explore the relativistic motion of the detector in Eq.~(\ref{trajectory}) how to affect the precision in the estimation of parameter $\beta=2\pi/a$ for the case of non-relativistic velocities, comparing with the results in Refs.~\cite{Tian2015,Wang2014} which shown that the uniformly accelerated detector $(w=0)$ moving along a spatially straight line degrade the QFI.
Similarly, substituting Eqs.~(\ref{final state}) and (\ref{Veffects}) into Eq.~(\ref{QFI}), we can also get the detailed formula of the QFI $F_\beta$, which does not contain any information about phase parameter $\phi$. It is needed to note that the expression is too long to exhibit here.

\begin{figure}[htbp]
\centering
\includegraphics[height=2in,width=3in]{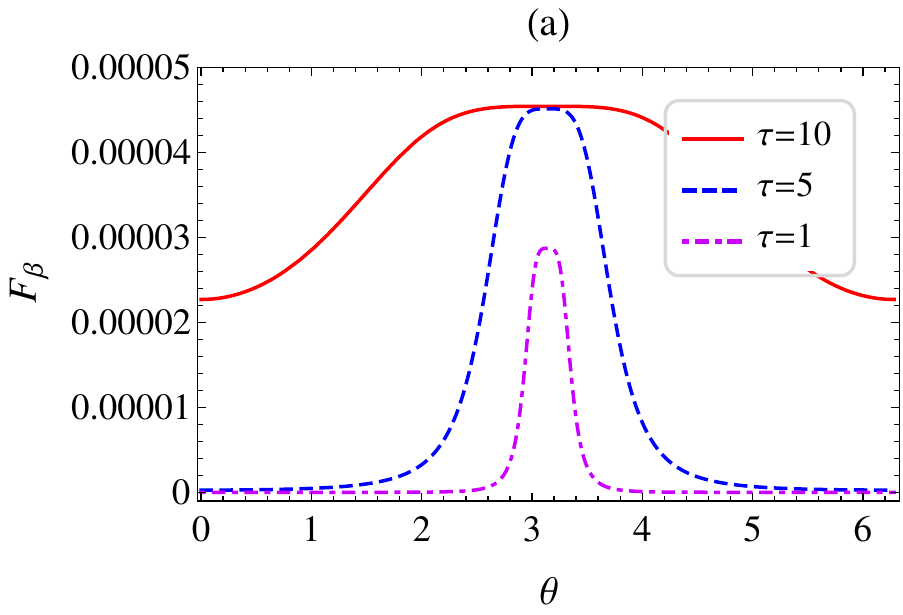}
\includegraphics[height=2in,width=3in]{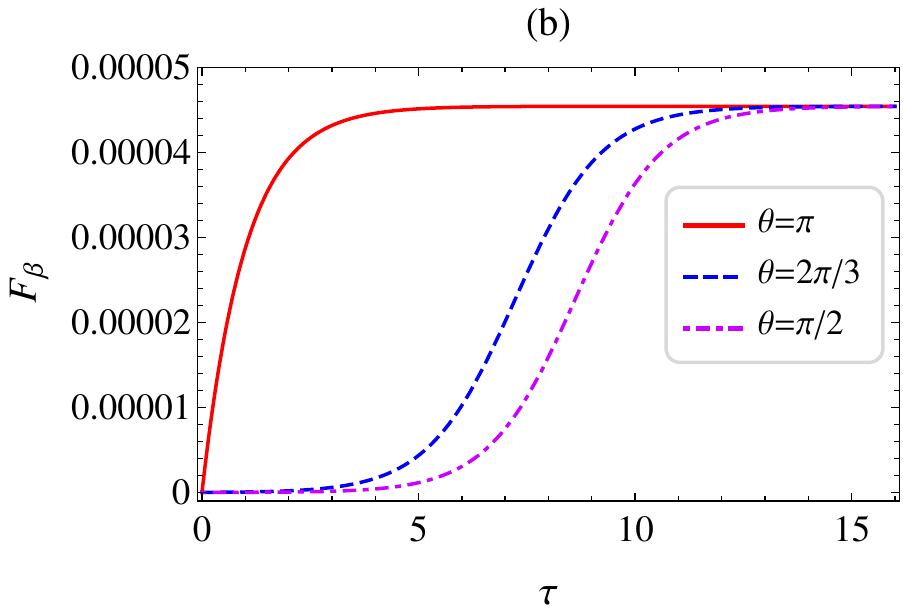}
\caption{ (a) The QFI $F_\beta$ as a function of the initial weight parameter $\theta$ with different effective time $\tau=10$ (solid line), $\tau=5$ (dashed line), $\tau=1$ (dot-dashed line); (b) The QFI  $F_\beta$ as a function of the effective time $\tau$ with $\theta=\pi$ (solid line), $\theta=2\pi/3$ (dashed line), $\theta=\pi/2$ (dot-dashed line). Here, we take the parameter $\beta=10$ and four-velocity component $w=0.01$.}\label{Q1}
\end{figure}
To clarify what value of initial weight parameter $\theta$ could allow better estimation, we plot, in Fig.~\ref{Q1}, the QFI $F_\beta$ as a function of the initial weight parameter $\theta$ (effective time $\tau$) with different effective time $\tau$ (initial weight parameter $\theta$). Here, we fix the parameter $\beta=10$ which was also considered in Ref.~\cite{Tian2015}.
It is interesting to note that in Fig.~3 of Ref.~\cite{Tian2015}, the behavior of QFI of $\beta$ for a uniformly accelerated detector $(w=0)$ is very similar with the behavior of QFI $F_\beta$ in Fig.~\ref{Q1} of this paper.
Therefore, we can deduce from the Fig.~\ref{Q1}~(a) that the QFI $F_\beta$ reaches the maximum value at $\theta=\pi$, which implies that the maximum sensitivity in the predictions for the parameter $\beta$ can be obtained by initially preparing the detector in its \emph{ground state}. Moreover, the symmetry of the QFI $F_\beta$ with respect to $\theta=\pi$ shows in Fig.~\ref{Q1}~(a). Besides, figure \ref{Q1}~(b) presents that the QFI $F_\beta$ is a monotonically increasing function of effective time $\tau$ during the initial period. However, when the detector evolves for a long enough time, i.e., $\tau\gg \frac{1}{A}$ with $1/A$ being the time scale for atomic transition, whatever the initial state of detector is prepared in, the QFI achieves the maximum and equals to each other, which represents that the optimal precision in the estimation of the parameter $\beta$ is completely unaffected by initial preparation of the detector if the effective time is long enough.

\begin{figure}[htbp]
\centering
\includegraphics[height=2in,width=3in]{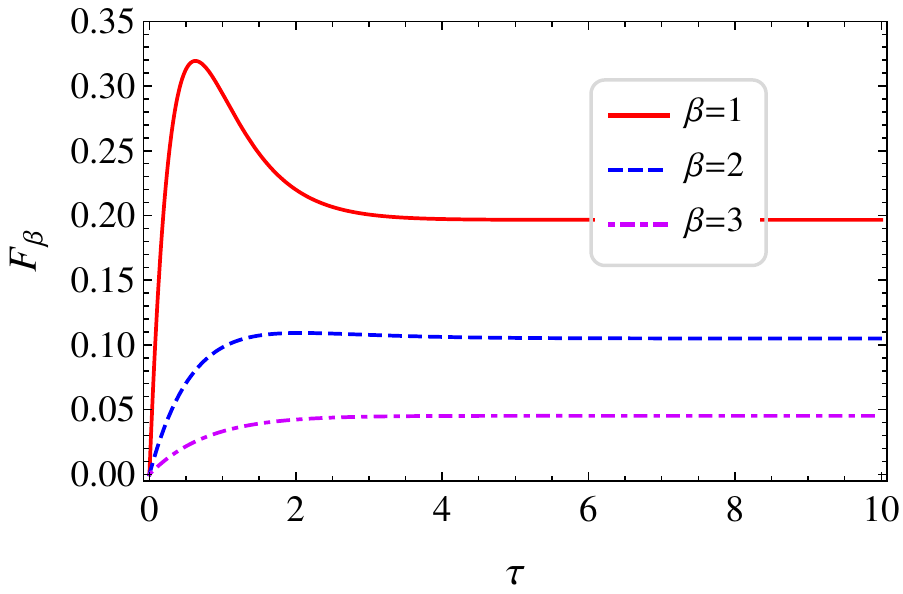}
\caption{ The QFI $F_\beta$ as a function of the effective time $\tau$ with different values of parameter $\beta$, i.e., $\beta=1,2,3$. By preparing the detector initially in the ground state $\theta=\pi$ and four-velocity component $w=0.01$. }\label{Q2}
\end{figure}
Furthermore, in Fig.~\ref{Q2}, we plot the QFI of parameter $\beta$ as a function of the effective time $\tau$ with different values of $\beta$ at fixed $w=0.01$ for $\theta=\pi$.  We find that the QFI $F_\beta$ saturates at different maximum values for different parameter $\beta$ in the limit of infinite effective time. However, we can see that when $\beta=1$, the QFI in the estimation of $\beta$ will increases for a while and starts to decrease but converges to nonzero value for a long enough time. This is due to the fact that for different values of parameter $\beta$, i.e., by taking different values of the effective four-acceleration of detector, the conditions of the detector eventually approaching to the equilibrium state are different. We can also obtain the same results of Ref.~\cite{Tian2015}, although it was not shown. This implies that for the small value $\beta=1$, the optimal precision in the estimation of $\beta$ can be obtained when the detector evolves for a finite time. Besides,
we note that the smaller the parameter $\beta$, i.e., for larger value of the effective four-acceleration parameter $a$, the higher the QFI $F_\beta$ is. That is, the highest precision in the estimation of parameter $\beta$ can be obtained for a larger four-acceleration parameter.

\begin{figure}[htbp]
\centering
\includegraphics[height=2in,width=3in]{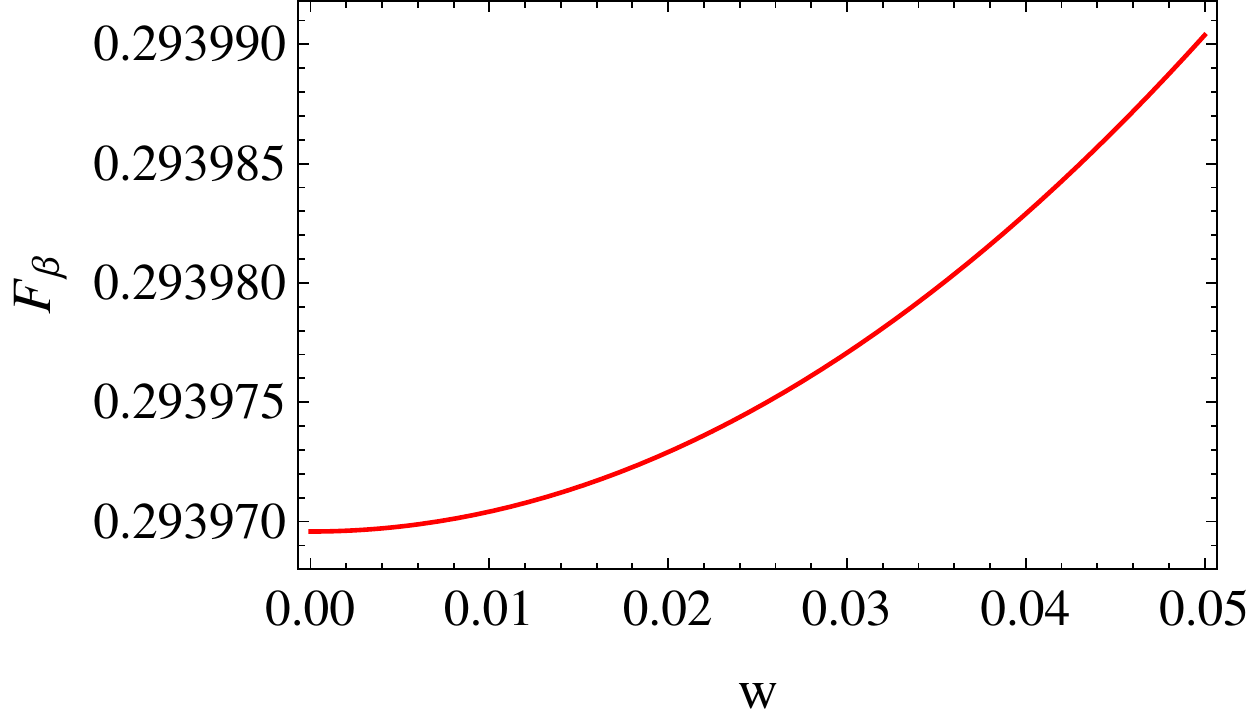}
\caption{ The QFI $F_\beta$ as a function of the four-velocity component $w$. Here, we take the detector in the ground state $\theta=\pi$, the effective time $\tau=1$ and the parameter $\beta=1$. }\label{Q3}
\end{figure}
Similarly, to analyze the four-velocity component for the accelerated two-level detector in the non-relativistic limit how to affect the precision in the estimation of parameter $\beta$, we plot the QFI $F_\beta$ as a function of the four-velocity component $w$ in Fig.~\ref{Q3}. We find that the higher the four-velocity component, the bigger the QFI is, i.e., the easier it is to achieve a given precision in the estimation of parameter $\beta$.
In this respect, comparing the above analysis with the results in Refs.~\cite{Tian2015,Wang2014}, we find that the relativistic motion of detector moving along such trajectory shown in Eq.~(\ref{trajectory}) can inhibition the degradation of the QFI, which implies that the precision in the estimation of parameter $\beta$ can be enhanced.

%%%%%%%%%%%%%%%%%%%%%%%%%%%%%%%%%%%%%%%%%%%%%%%%%%%%%%%%%%%%%%%%%%%%%%%%%%%%%%

\subsection{In the case for ultra-relativistic velocities} \label{sectionIIIB}
Now we consider the detector moving along the trajectory in Eq.~(\ref{trajectory}) for the case of ultra-relativistic velocities, i.e., $w\rightarrow\infty$, the two-point correlation function for the massless scalar field in Eq.~(\ref{correlation}) which is suppressed as
\begin{eqnarray}\label{ultrarelativistic}
G^+(x-x')=-\frac{a^2}{16\pi^2}\bigg[\sinh^2\bigg(\frac{a\Delta\tau}{2}-i\epsilon a\bigg)\bigg]^{-1}\frac{1}{w^4}.
\end{eqnarray}
According to Eqs.~(\ref{gamma}) and (\ref{ultrarelativistic}), and by invoking the contour integral, the A and B in Eq.~(\ref{final state}) can be obtained as
\begin{eqnarray}\label{ultraAB}
A=\gamma_0\bigg(\frac{e^{2\pi\omega_0/a}+1}{e^{2\pi\omega_0/a}-1}\bigg)\frac{1}{w^4},\;\;\;B=-\gamma_0\frac{1}{w^4}.
\end{eqnarray}
Thus, when $w\rightarrow\infty$, the A and B in Eq.~(\ref{ultraAB}) for ultra-relativistic velocities are given by
\begin{eqnarray}\label{AB}
A\rightarrow0,\;\;\;B\rightarrow0.
\end{eqnarray}
This suggests that the detector evolves with time as a closed system, whose evolution is completely unaffected by the external environment and detector's motion.
Because the trajectory of detector in Eq.~(\ref{trajectory}) is modified by the factor $1/\sqrt{1+w^2}$, such trajectory becomes constant in the ultra-relativistic limit.
Therefore, submitting Eq.~(\ref{AB}) into Eq.~(\ref{final state}), the Bloch vector of the state $\rho(\tau)$ with respect to the proper time $\tau$ can easily be written as
\begin{eqnarray}\label{final state1}
\omega_1(\tau)&=&\sin\theta \cos(\omega_0\tau+\phi)\;,\nonumber\\
\omega_2(\tau)&=&\sin\theta \sin(\omega_0\tau+\phi) \;,\nonumber\\
\omega_3(\tau)&=&\cos\theta \;.
\end{eqnarray}
With the help of Eqs.~(\ref{QFI}) and (\ref{final state1}), one can obtain the QFI in terms of $\theta$ and $\phi$ as
\begin{eqnarray}\label{QFIultrare}
F_{\theta}=1,\;\;\;F_\phi=\sin^2\theta,
\end{eqnarray}
which shows that the QFI is time independent.

Note that, in contrast to the previous results show that in the non-relativistic limit the QFI will be enhanced with the increase of the four-velocity component $w$, the QFI in Eq.~(\ref{QFIultrare}), for the case of ultra-relativistic velocities, will be never subjected to affected by the environment and remains constant with time, as if the detector were a closed system. More interestingly, it is worth emphasizing that the Unruh-DeWitt detector moving along such trajectory in the ultra-relativistic limit, has the same impact as the results of that a system interacted with environment by the presence of boundaries in certain circumstances~\cite{Jin2015,1Liu2016,2Liu2016,Yang2018,Yang2019}, which indicates that the QFI can be shield from the influence of the environment.

It is worth emphasizing that when the detector is in inertial motion, i.e., when the detector moving with a single constant velocity, although the QFI does not dependent on the detector's velocity, it is still degraded exponentially with the evolution time as a result of the interaction between the detector and field.
Therefore, the precision of
the parameters estimation is decreased.
Moreover, for an uniformly accelerated detector, the decrease of QFI over time would be enhanced by the acceleration, as shown in
Ref. \cite{Yang2018}. However, when the detector moves with a combination of the linear accelerated motion and a component of the four-velocity $w=dy/d\tau$, we find the QFI depends on both the velocity and acceleration, and the velocity will suppress the
degradation of QFI compared with
the individual acceleration case in the non-relativistic limit. What is more, in the ultra-relativistic limit, the QFI may be shielded from the effects of the detector's motion, and even remains constant with time as if it were a closed system.
This intriguing behaviors come from the composite effect of both velocity and acceleration, and are not valid for the individual accelerated case
and individual constant-velocity case.

%%%%%%%%%%%%%%%%%%%%%%%%%%%%%%%%%%%%%%%%%%%%%%%%%%%%%%%%%%%%%%%%%%%%%%%%%%%%%%
\section{conclusions} \label{sectionIV}
In the framework of open quantum systems, we studied the dynamics of the QFI of the parameters estimation for a detector interacted with massless scalar field.
For the detector moving along a spatially straight line with a constant velocity, we found that the QFI always will be degraded by the external environment, but unaffected by the velocity.
Besides, for the uniformly linear accelerated detector ($w=0$), the acceleration will cause the QFI and thus the precision limit of parameter estimation to degrade, as shown in Ref.~\cite{Yang2018}.

However, when the detector moving along an unbounded spatial trajectory in a two-dimensional spatial plane with constant independent magnitudes of both the four-acceleration $a$ and also having a component of four-velocity $w=dy/d\tau$ constant, the QFI of this detector in two different situations, in the non-relativistic and ultra-relativistic limit, have been detailedly considered.
In the non-relativistic limit, we can achieve the optimal strategy for the parameters estimation by preparing the proper probe and adjusting the interaction parameters. Moreover, the four-velocity component will suppress the degradation of the QFI. That is, the precision of the parameters estimation can be enhanced by the relativistic motion of the detector following trajectory (\ref{trajectory}) for non-relativistic velocities.
In the ultra-relativistic limit, counterintuitively,  the QFI remains constant with time due to the relativity correction to the four-velocity component, which implies that the precision of the parameters estimation can be completely unaffected by the the external environment and detector's motion, as if the detector were a closed system.

%%%%%%%%%%%%%%%%%%%%%%%%%%%%%%%
\begin{acknowledgments}
This work was supported by the National Natural Science Foundation
of China under Grant Nos. 12065016, 11875025 and 11905218; Hunan Provincial Natural Science Foundation of China under Grant No. 2018JJ1016; The CAS Key Laboratory for Research in Galaxies and Cosmology, Chinese Academy of Science (No. 18010203); The Physics Key Discipline of Liupanshui Normal University under Grant No. LPSSYZDXK201801; The cultivation project of Master's degree of Liupanshui Normal University under Grant No. LPSSYSSDPY201704. X. Liu thanks for the Young scientific talents growth project of the department of education of the department of education of Guizhou province under Grant No. QJHKYZ[2019]129; The talent recruitment program of Liupanshui normal university of China under Grant No. LPSSYKYJJ201906.

\end{acknowledgments}

\end{document}